\documentstyle[prd,aps,preprint,epsfig,floats]{revtex}
\tightenlines

\def\dsodt{ds_{23}\over dt}
\def\dstdt{ds_{13}\over dt}
\def\dsthdt{ds_{12}\over dt}

\def\be{\begin{equation}}
\def\ee{\end{equation}}
\def\ba{\begin{eqnarray}}
\def\ea{\end{eqnarray}}
\def\br{\begin{array}}
\def\er{\end{array}}

\def\DESepsf(#1 width #2){\epsfxsize=#2 \epsfbox{#1}}

\begin{document}

\draft


\title{{\Large\bf Seesaw Mechanism and Its Implications}}
\author{\bf  R.N. Mohapatra}

\address{ Department of Physics, University of Maryland, College Park,
MD-20742, USA}
\maketitle

\begin{abstract}
 The seesaw mechanism is introduced and some of its different
realizations and applications are
discussed. It is pointed out how they can be used to understand the
bi-large mixing patterns among neutrinos in combination with the
assumptions about high scale physics such as grand unification or
quasi-degeneracy of neutrino masses.
\end{abstract}

\vskip1.0in

\section{Introduction}
The discovery of neutrino masses and mixings has been an important
milestone in
the history of particle physics and rightly qualifies as the
first evidence for new physics beyond the standard model. The amount of
new information on neutrinos already established from various neutrino
oscillation searches has provided very strong clues to new symmetries
of particles and new directions for unification. Enough puzzles have
emerged making this field a hotbed for theory
research with implications ranging all the way from supersymmetry and
grand unification to cosmology and astrophysics.

A major cornerstone for the theory research in this field has
been the seesaw mechanism introduced 25 years ago in four
independently written papers\cite{seesaw} to understand why
neutrino masses are so much smaller than the masses of other
fermions of the standard model. Even though there was no solid
evidence for neutrino masses then, there were very well motivated
extensions of the standard models that led to nonzero masses for
neutrinos. It was therefore incumbent on those models that they
have a mechanism for understanding why upper limits on neutrino
masses known at that time were so small and the seesaw mechanism
was introduced in the context of specific such models in the year 1979 
e.g. horizontal, left-right and SO(10) models to achieve this
goal. A general operator description of small neutrino mass without any
specific model was written down the same year\cite{weinberg}. A very
minimal nonsupersymmetric SO(10) model was constructed soon after as an
application\cite{witten}. It was clear from this early enthusiasm about
the idea that if the experimental evidence for neutrino masses ever
appeared then, seesaw mechanism would be a major tool in understanding
its various ramifications. As we see below, this has indeed turned out to
be the case.

\section{Seesaw mechanism}
To appreciate the simplicity and beauty of the seesaw mechanism,
let us start with a discussion of neutrino mass in the standard model.
It is based on the gauge group
$SU(3)_c\times SU(2)_L\times U(1)_Y$ group under which the quarks and
leptons transform as follows: Quarks ${\bf Q}^T_L\equiv (u_L,d_L)(3,2,
{1\over
3})$; ${\bf u}_R (3, 1, {4\over 3})$ ; ${\bf d}_R(3, 1,-\frac{2}{3})$;
leptons ${\bf L}^T\equiv (\nu_L,e_L)(1, 2, -1)$;  ${\bf e}_R (1,1,-2)$;
Higgs Boson ${\bf H}(1, 2, +1)$; Color Gauge Fields  ${\bf G}_a(8, 1, 0)$;
Weak Gauge Fields  ${\bf W^{\pm}, Z, \gamma} (1,3+1,0)$.

The electroweak symmetry $SU(2)_L\times U(1)_Y$ is broken by the vacuum
expectation of the Higgs doublet $<H^0>=v_{wk}\simeq 246$ GeV, which gives
mass to the gauge bosons and the fermions, all fermions except the
neutrino. The model had been a complete success in describing all
known low energy phenomena, until the evidence for neutrino masses
appeared.

Note that there is no right handed neutrino in
the standard model and this directly leads to the fact that neutrinos are
massless  at the tree level. This result holds not only to
all orders in
perturbation theory but also when nonperturbative effects are taken into
account due to the existence of an exact B-L symmetry of the standard
model. It would therefore appear that nonzero neutrino mass ought to be
connected to breaking of B-L symmetry.

A simple way to generate neutrino masses is to introduce right handed
neutrinos $N_R$, one per family into the standard model. The standard
model Lagrangian now allows for a new
Yukawa coupling of the form $h_\nu\bar{L}HN_R$ which after electroweak
symmetry breaking leads to a neutrino mass $\sim h_\nu v_{wk}$. Since
$h_\nu$ is expected to be of same order as the charged fermion couplings
in the model, this mass is much too large to describe neutrino
oscillations. Luckily, since the
$N_R$'s are singlets under the standard model gauge group,
they are allowed to have Majorana masses unlike the
charged fermions.  We denote them by $M_RN^T_RC^{-1}N_R$ (where $C$
is the Dirac charge conjugation matrix). The masses $M_R$
are not constrained by the gauge symmetry
and can therefore be arbitrarily large (i.e. $M_R \gg h_\nu v_{wk}$). This
together with mass induced 
by Yukawa couplings (called the Dirac mass) leads to a the mass matrix for
the neutrinos (left and
right handed neutrinos together) which has the form
\begin{eqnarray}
{ M}_\nu~=~\left(\begin{array}{cc}o & M_D\\ M^T_D & M_R\end{array}\right)
\end{eqnarray}
where $M_D$ and $M_R$ are $3\times 3$ matrices. Diagonalizing this mass
matrix, one gets the mass matrix for the light neutrino masses to be as
follows:
\begin{eqnarray}
{ M}_\nu~=~-M^T_DM^{-1}_RM_D
\end{eqnarray}
Since as already noted $M_R$ can be much larger than $M_D$ which is likely
to be of order $=~h_\nu v_{wk}$, one finds
that $m_\nu \ll m_{e,u,d}$ very naturally. This is known as the seesaw
mechanism\cite{seesaw} and it provides a natural explanation of why
neutrino masses are small.

Seesaw mechanism of course raises its own questions:

\begin{itemize}

\item what is the scale of $M_R$ and what determines it ?

\item  Is there a natural reason for the existence of the right handed
neutrinos ? 

\item  Is the seesaw mechanism by itself enough to explain all aspects of
neutrino masses and mixings.

\end{itemize}

Below, we try to answer some of these questions and discuss how far one
can go towards explaining observations.

 \section{Why seesaw mechanism is so appealing ?}
Even though the standard model has enjoyed incredible success in
explaining all low energy observations, it has long
been recognized that it cannot be a complete theory due to many issues
it leaves unaddressed, e.g. the gauge hierarchy problem, strong CP problem
as well as the fermion mass and mixing problem etc. Besides, it also has
some aesthetic inadequacies that cry out for new physics. In the latter
category are such
questions as (i) an omission of the right handed neutrino which
simple quark lepton symmetry observed in weak interaction would have
demanded; (ii) a somewhat adhoc definition of the electric charge in terms
of an U(1) charge called ``weak hypercharge $Y$'' i.e. $Q~=~I_{3L} +
\frac{Y}{2}$ and (iii) the origin of parity violation. It turns out that
inclusion of the three right handed neutrinos helps to remove the
aesthetic cloud that hangs over the standard model while maintaining all
its phenomenological successes.

The fact that the addition of one right handed neutrino per generation to
the standard model restores quark lepton symmetry is obvious. But in 1973,
Pati and Salam pointed out that there is a compelling symmetry reason for
including the right handed neutrino if one considers leptons as a fourth
color in Nature\cite{ps}, the other three being associated with quarks.
In fact in the presence of the $N_R$'s, the minimal anomaly free gauge
group of weak interactions expands beyond the standard model and
becomes the left-right symmetric group $SU(2)_L\times SU(2)_R\times
U(1)_{B-L}$\cite{lrs} which is a subgroup of the $SU(2)_L\times
SU(2)_R\times SU(4)_c$ group introduced by Pati and Salam.
 We will see below that both these gauge groups lead to a
picture of weak interaction which is fundamentally different from that
envisaged in the standard model in that {\it weak interactions like the
strong and
electromagnetic ones are parity conserving at very high energies 
and observed maximally parity violating V-A structure of low energy weak
processes is a consequence of gauge symmetry breaking. }

To see this explicitly, we consider the left-right symmetric theory of
weak interactions under which fermions and Higgs bosons transform as
follows:
${\bf Q}_L\equiv \left(\begin{array}{c}u_L \\ d_L\end{array}\right)$
(2,1,$+ {1 \over 3}$);
${\bf Q}_R \equiv \left(\begin{array}{c}u_R \\ d_R\end{array}\right)$
(1,2,$ {1 \over 3}$);
${\bf L}_L\equiv \left(\begin{array}{c}\nu_L \\ e_L\end{array}\right)$
 (2,1,$- 1$);
${\bf L}_R\equiv \left(\begin{array}{c}\nu_R \\ e_R\end{array}\right)$
 (1,2,$- 1$). The Higgs fields transform as
${\bf \phi}$ (2,2,0) ;
${\bf \Delta}_L$ (3,1,+ 2);
${\bf \Delta}_R$ (1,3,+ 2).

It is clear that this theory leads to a weak interaction Lagrangian of the
form
\begin{eqnarray}
{ L}_{wk}~=~ \frac{g}{2}\left(\vec{j}^\mu_L\cdot \vec{W}_{L,\mu}
+\vec{j}^\mu_R\cdot \vec{W}_{R,\mu}\right)
\end{eqnarray}
which is parity conserving prior to symmetry breaking.
 Furthermore, in this theory, the elctric
charge formula is given by\cite{marshak}:
\begin{eqnarray}
Q~=~I_{3L}~+~I_{3R}~+~\frac{B-L}{2},
\end{eqnarray} 
where each term has a physical meaning unlike the 
case of the standard model. When only the
gauge symmetry $SU(2)_R\times U(1)_{B-L}$ is broken down,
 one finds the relation $\Delta
I_{3R}~=-\Delta\left(\frac{B-L}{2}\right)$. This connects $B-L$ breaking
i.e.$\Delta(B-L)\neq 0$
 to the breakdown of parity symmetry i.e. $\Delta I_{3R}\neq 0$.
It also reveals the true meaning of the standard model hypercharge as
$\frac{Y}{2}~=~I_{3,R}~+\frac{B-L}{2}$.

To discuss the implications of these observations, note that in stage I,
the gauge symmetry is broken by the Higgs
multiplets $\Delta_L(3,1,2)\oplus \Delta_R(1,3,2)$ to the standard
model and in stage II by the bidoublet $\phi(2,2,0)$ as in the standard
model.
In the first stage, the right
handed neutrino picks up a mass of order $f<\Delta^0_R>\equiv
fv_R$. Denoting the left and right handed neutrino by $(\nu, N)$ (in a
two component notation), the mass
matrix for neutrinos at this stage looks like
\begin{eqnarray}
{ M}^0_\nu~=~\left(\begin{array}{cc} 0 & 0 \\ 0 & fv_R\end{array}\right)
\end{eqnarray} 
At this stage, familiar standard model particles are all massless.
As soon as the standard model symmetry is broken by the bidoublet $\phi$
i.e. $<\phi>\equiv diag(\kappa, \kappa')$, the W and Z
boson
as well as the fermions pick up mass. I will generically denote
$\kappa,\kappa'$ by a common symbol $v_{wk}$. The contribution to neutrino
mass at this stage look like
\begin{eqnarray}
{ M}^0_\nu~=~\left(\begin{array}{cc} fv_L & h v_{wk} \\ hv_{wk} &
fv_R\end{array}\right)
\end{eqnarray}
where $v_L~=~\frac{v^2_{wk}}{v_R}$. The appearance of the $fv_L$ term is a
reflection of parity
invariance of the model. Note that except for the $\nu\nu$
entry, the neutrino mass matrix in Eq.(6) is exactly in the same form as
in Eq. (1). Diagonalizing this
matrix, we get a modified seesaw formula for the light neutrino
mass matrix
\begin{eqnarray}
{ M}_\nu~=~fv_L-h^T_\nu f^{-1}_Rh_\nu
\left(\frac{v^2_{wk}}{v_R}\right)
\end{eqnarray}
 The important point to note is that $v_L$ is 
suppressed by the same factor as the second term in Eq. (7) so that
despite the new
contribution to neutrino masses, seesaw suppression
remains\cite{seesaw2}. This is called the type II seesaw whereas
the formula in Eq. (1) is called type I seesaw formula\footnote{The 
triplet contribution to neutrino masses without the seesaw suppression
was considered in \cite{valle} and triplet contribution by itself
with seesaw suppression outside the framework of parity symmetric models
have been considered in \cite{sarkar}. In the limit of large right
handed neutrino masses, type II seesaw formula for neutrino masses
reduces to the triplet seesaw formula.}.

An important physical meaning of the seesaw formula is brought out when
it is viewed in the context of left-right models. Note that
$m_\nu\rightarrow 0$ when $v_R$ goes to infinity. In the same
limit the weak interactions become pure V-A type. Therefore, left-right
model derivation of the seesaw formula smoothly connects smallness of
neutrino mass with suppression of V+A part of the weak interactions
providing an important clarification of a major puzzle of the standard
model i.e. why are weak interactions are near maximally parity violating ?
The answer is that they are near maximally parity violating because the
neutrino mass happens to be small. This point was emphasized in the fourth
paper in Ref. [1].

In a subsequent section, we will discuss the connection of the seesaw mass
scale with the scale of grand unification, which is suggested by the value
of the atmospheric $\Delta m^2_A$. SO(10) is the simplest gauge group that
contains the right handed neutrino needed to implement the seesaw
mechanism and also it is important to note that the left-right
symmetric gauge group is a subgroup of the SO(10) group, which therefore
provides an attractive over all grand unified framework for the
discussion of neutrino masses. The extra bonus one may expect is that
since bigger symmetries tend to relate different parameters of a theory,
one may be able to predict neutrino masses and mixings. We will present a
model where indeed this happens.

We also note that since type I seesaw involves the Dirac mass of
the
neutrino, which is likely to scale with generation the same
way as the charged fermions of the standard model, unless
there is extreme hierarchy among the right handed neutrinos, one would
expect the $\nu$ spectrum to be hierarchical. On the other hand, it has
been
realized for a long time\cite{caldwell} that if neutrino masses are
quasi-degenerate, it is a tell-tale sign of type II seesaw with the
triplet vev term being the dominant one. However, a normal hierarchy can
also arise with type II seesaw as we discuss in the example
below. 

 \section{Seesaw and large neutrino mixings}
While seesaw mechanism provides a simple framework for understanding the
smallness of neutrino masses, it does not throw any light on 
the question of why neutrino mixings are large. The point is that 
mixings are a consequence of the structure of the light neutrino mass
matrix and the seesaw mechanism is only statement about the scale of new
physics. This can also be understood by doing a simple parameter
counting. If we work in a basis where the right handed neutrino masses are
diagonal, there are 18 parameters describing the seesaw formula for
neutrino masses - three RH neutrino masses and 15 parameters in the Dirac
mass matrix. On the other hand, there are only nine observables (three
masses, three mixing angle snd three phases) describing low energy
neutrino sector. Thus there are twice as many parameters as observables.
As a result, the neutrino mass matrix needs inputs beyond the simple
seesaw mechanism to fix the neutrino mass matrix.

In order to understand large mixings, one has to go beyond the simple
seesaw mechanism to particular models. This is any way necessary to limit
the scale of the right handed neutrino far below the Planck scale as seems
to be the case. Many such models have
been considered that use horizontal symmetry, grand unification, discrete
symmetries, assumption of single right handed neutrino dominance
etc.\cite{king} to derive large mixings. In the following section, I
will focus on a recently
discussed minimal SO(10) model, where without any assumption other than
SO(10) grand unification, one can indeed predict all but one neutrino
parameters. I will then consider a case where assumption of
quasi-degeneracy in the neutrino spectrum at high scale leads in a natural
way via radiative corrections to large mixings at low energies.

To understand the fundamental physics behind neutrino mixings, we first
write down the neutrino mass matrix that leads to maximal solar and
atmospheric mixing. We consider the case of normal hierarchy where we have
\begin{eqnarray}
{ M}_\nu~=~\frac{\sqrt{\Delta
m^2_A}}{2}\left(\begin{array}{ccc}c\epsilon
&b\epsilon &d\epsilon\\ b\epsilon & 1+a\epsilon & -1 \\
d\epsilon & -1 & 1+\epsilon\end{array}\right)
\end{eqnarray}
where $\epsilon \simeq \sqrt{\frac{\Delta m^2_\odot}{\Delta m^2_A}}$ and
parameters $a,b,c,d$ are of order one. Any theory of neutrino which
attempts to explain the observed mixing pattern for the case of normal
hierarchy must strive to get a mass matrix of this form. In the next
section, we give a simple example of a minimal SO(10) grand unified theory
that gives this mass matrix without any extra assumptions.

\section{A predictive minimal SO(10) theory for neutrinos}
 
The main reason for considering SO(10) for neutrino masses is that its
{\bf 16} dimensional spinor representation consists of all fifteen
standard model fermions plus the right handed neutrino arranged according
to the it $SU(2)_L\times SU(2)_R\times SU(4)_c$ subgroup\cite{ps} as
follows:
\begin{eqnarray}
{\bf \Psi}~=~\left(\begin{array}{cccc}u_1 & u_2 & u_3 & \nu\\ d_1 & d_2
& d_3& e\end{array}\right)
\end{eqnarray} 
There are three such spinors for three fermion families. 

In order to implement the seesaw mechanism in the SO(10) model, one must
break the B-L symmetry, since the right handed neutrino mass breaks this
symmetry. One implication of this is that the seesaw scale is at or
below the GUT scale. Secondly in the context of supersymmetric
SO(10) models, the way B-L breaks has profound consequences for
low energy physics. For instance, if B-L is broken by a Higgs field
belonging to the {\bf 16} dimensional Higgs field (to be denoted by
$\Psi_H$), then the field that
acquires a nonzero vev has the quantum numbers of the $\nu_R$ field
i.e. B-L breaks by one unit. In this case higher dimensional operators
of the form $\Psi\Psi\Psi\Psi_H$ will lead to R-parity violating operators
in the effective low energy MSSM theory such as $QLd^c, u^cd^cd^c$ etc
which can lead to large breaking of lepton and baryon number symmetry and
hence unacceptable rates for proton decay. This theory also has no dark
matter candidate without making additional assumptions.

On the other hand, one may break B-L by a {\bf 126} dimensional Higgs
field. The member of this multiplet that acquires vev has $B-L=2$ and
leaves R-parity as an automatic symmetry of the low energy Lagrangian.
 There is a naturally stable dark matter in this case.
It has recently been shown that this class of models lead to a very
predictive scenario for neutrino mixings\cite{babu,last,goran,goh}. We
summarize this model below.

As already noted earlier, any theory with asymptotic parity symmetry
leads to type II seesaw formula. It turns out that if the B-L symmetry is
broken by {\bf 16} Higgs fields, the first term in the type II seesaw
(effective triplet vev induced term) becomes very small compared to the
type I term. On the other hand, if B-L is broken by a {\bf 126} field,
then the first term in the type II seesaw formula is not necessarily small
and can in principle dominate in the seesaw formula. We will discuss a
model of this type below.

The basic ingredients of this model are that one considers only two Higgs
multiplets that contribute to fermion masses i.e. one
{\bf 10} and
one {\bf 126}. A unique property of the {\bf 126}
multiplet is that it not only breaks the B-L symmetry and therefore
contributes to
right handed neutrino masses, but it also contributes to charged fermion
masses by virtue of the fact that it contains MSSM doublets which mix with
those from the {\bf 10} dimensional multiplets and survive down to the
MSSM scale. This leads to a tremendous reduction of  the number of
arbitrary parameters, as we will see below.

There are only two Yukawa coupling matrices in this model: (i) $h$ for
the {\bf 10} Higgs and (ii) $f$ for the {\bf 126} Higgs.
SO(10) has the property that the Yukawa couplings involving the {\bf 10}
and {\bf 126} Higgs representations are symmetric. Therefore
if we assume that CP violation arises from other sectors of the theory
(e.g. squark masses) and work in a basis where one of these two sets
of Yukawa coupling matrices is diagonal, then it will have
only nine parameters. Noting the fact that the (2,2,15) submultiplet of
{\bf 126} has a pair of standard model doublets that contributes to
charged fermion masses, one can write the quark and lepton mass matrices
as follows\cite{babu}:
\begin{eqnarray}
M_u~=~ h \kappa_u + f v_u \\\nonumber
M_d~=~ h \kappa_d + f v_d \\  \nonumber
M_\ell~=~ h \kappa_d -3 f v_d \\  \nonumber
M_{\nu_D}~=~ h \kappa_u -3 f v_u \\\nonumber
\end{eqnarray}
where $\kappa_{u,d}$ are the vev's of the up and down standard model
type Higgs fields in the {\bf 10} multiplet and $v_{u,d}$ are the
corresponding vevs for the same doublets in {\bf 126}.
Note that there are 13 parameters in the above equations and there are 13
inputs (six quark masses, three lepton masses and three quark mixing
angles and weak scale). Thus all parameters of the model that go into
fermion masses are determined.

 To determine the light neutrino masses, we use the seesaw
formula in Eq. (7), i.e.
\begin{eqnarray}
{ M}_\nu~=~fv_L-h^T_\nu f^{-1}_Rh_\nu
\left(\frac{v^2_{wk}}{v_R}\right).
\end{eqnarray}
The coupling matrix ${f}$ is nothing but the {\bf 126}
Yukawa coupling that appears in Eq. (10). Thus all parameters that give
neutrino mixings except an overall scale are determined. These models were
extensively discussed in
the last decade\cite{last} using type I seesaw formula. Their predictions
for neutrino masses and mixings are either ruled out or are at best
marginal.

There has been a revival of these models due to an observation for a two
generation version of it\cite{goran}. It was pointed out in
Ref.[14] that if the
direct triplet term in type II seesaw dominates, then it provides a very
natural understanding of the large atmospheric mixing angle for the case
of two generations without invoking any symmetries. Subsequently it was
shown\cite{goh} that the same $b-\tau$ mass convergence also 
provides an explanation of large solar mixing as well as small
$\theta_{13}$ making the model realistic and experimentally interesting.

 A simple way to see how large mixings arise in this model is to note that
when the triplet term dominates the seesaw
formula,  we have the neutrino mass matrix ${ M}_\nu \propto f$,
where $f$ matrix is the {\bf 126} coupling to fermions discussed earlier.
Using the above equations, one can derive the following
sumrule (sumrule was already noted in the second reference of
\cite{last}):
\begin{eqnarray}
{ M}_\nu~=~ c (M_d - M_\ell)
\label{key}
\end{eqnarray}
where numerically $c\approx 10^{-9}$ GeV.
To see how this leads to large atmospheric and solar mixing,
let us work in the basis where the down quark mass matrix is diagonal. All
the quark mixing effects are then in the up quark mass matrix i.e.
$M_u~=~U^T_{CKM}M^d_u U_{CKM}$. Note further that the minimality of the
Higgs content leads to the following sumrule among the mass matrices:
\begin{eqnarray}
k \tilde{M}_{\ell}~=~r\tilde{ M}_d +\tilde{ M}_u
\end{eqnarray}
where the tilde denotes the fact that we have made the mass matrices
dimensionless
by dividing them by the heaviest mass of the species i.e. up quark mass
matrix by $m_t$, down quark mass matrix by $m_b$ etc. $k,r$ are functions
of the symmetry breaking parameters of the model.
Using the Wolfenstein parameterization for quark mixings, we can conclude
that that we have
\begin{eqnarray}
M_{d,\ell}~\approx ~m_{b,\tau}\left(\begin{array}{ccc}\lambda^3 &
\lambda^3
&\lambda^3\\ \lambda^3 & \lambda^2& \lambda^2 \\ \lambda^3 & \lambda^2 &
1\end{array}\right)
\end{eqnarray}
where $\lambda \sim 0.22$ and the matrix elements are supposed to give
only the approximate order of magnitude. 

An important consequence of the relation between the
charged lepton and the quark mass matrices in Eq. (13) is that the charged
lepton
contribution to the neutrino mixing matrix i.e. $U_\ell \simeq {\bf 1} +
O(\lambda)$ or close to identity matrix. As a result
the neutrino mixing matrix
is given by $U_{PMNS}~=~U^{\dagger}_\ell U_\nu \simeq U_\nu$, since in
$U_\ell$, all mixing angles are small. Thus the dominant
contribution to large mixings will come from $U_\nu$, which in turn
will be dictated by the sum rule in Eq. (12). Let us now see how how this
comes about.

As we extrapolate the quark
masses to the GUT scale, due to the fact that $m_b-m_\tau \approx
m_{\tau}\lambda^2$ for a wide range of values of tan$\beta$, the
neutrino mass matrix
$M_\nu~=c(M_d-M_\ell)$ takes roughly the form
\begin{eqnarray}
M_{\nu}~=c(M_d-M_\ell)\approx ~m_0\left(\begin{array}{ccc}\lambda^3 &
\lambda^3
&\lambda^3\\ \lambda^3 & \lambda^2 & \lambda^2 \\ \lambda^3 & \lambda^2
& \lambda^2\end{array}\right)
\end{eqnarray}
This mass matrix is in the form discussed in Eq. (8) and 
it is easy to see that both the $\theta_{12}$
(solar angle) and $\theta_{23}$ (the atmospheric angle) are now large. The
detailed magnitudes of these angles of course depend on the details of the
quark masses at the GUT scale. Using the extrapolated values of the quark
masses and mixing angles to the GUT scale, the
predictions of this model for various oscillation parameters are given in
Ref.[15]. The predictions for the
solar and atmospheric mixing angles fall within 3 $\sigma$ range of the
present central values. Specifically the
prediction for $U_{e3}$ (see Fig. 1) can be tested in MINOS as well as
other planned Long Base Line neutrino experiments such as Numi-Off-Axis,
JPARC etc. 
\begin{figure}
\begin{center}
\epsfxsize8cm\epsffile{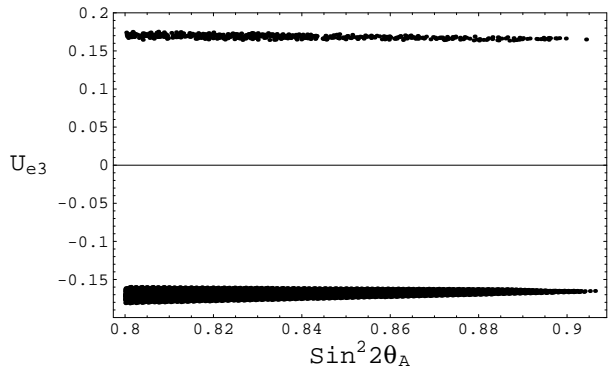}
\caption{
The figure shows the predictions of the minimal SO(10) model for
$sin^22\theta_{A}$
and $U_{e3}$ for the allowed range of parameters in the model. Note that
$U_{e3}$ is very close to the upper limit allowed by the existing reactor
experiments.
\label{fig:cstr3}}
\end{center}
\end{figure}

There is a simple explanation of why the $U_{e3}$ comes out to be
large. This can also be seen from the mass sumrule in
Eq.\ref{key}. Roughly, for a matrix with hierarchical eigen values as is
the case here, the mixing angle $tan 2\theta_{13}\sim
\frac{M_{\nu,13}}{M_{\nu,33}}\simeq \frac{\lambda^3
m_\tau}{m_b(M_U)-m_\tau(M_U)}$. Since to get large mixings, we need
$m_b(M_U)-m_\tau(M_U)\simeq m_\tau \lambda^2$, we see that $U_{e3}\simeq
\lambda$ upto a factor of order one. Indeed the detailed calculations lead
to $0.16$ which is not far from this value.

\section{CP violation in the minimal SO(10) model}
In the discussion given above, it was assumed that CP violation is non-CKM
type and resides in the soft SUSY breaking terms of the Lagrangian. The
overwhelming evidence from experiments seem to be that CP violation is
perhaps is of CKM type. It has
recently been pointed out that with slight modification, one can include
CKM CP violation in the model\cite{mimura}. The basic idea is to include
all higher dimensional operators of type $h'{\Psi \Psi}
\bar{\Delta}{\Sigma}/M$ where $\bar{\Delta}$ and $\Sigma$ denote
respectively the {\bf 126} and the {\bf 210} dimensional representation.
It is then clear that those operators transforming as {\bf 10} and {\bf
126} representations will simply redefine the $h,f$ coupling matrices and
add no new physics. On the other hand the higher dimensional operator that
transforms like an effective {\bf 120} representation will add a new piece
to all fermion masses. Now suppose we introduce a parity symmetry into the
theory which transforms $\Psi$ to ${\Psi^c}^*$, then it turns out that the
couplings $h$ and $f$ become real and symmetric matrices whereas the {\bf
120} coupling (denoted by $h'$) becomes imaginary and antisymmetric. This
process introduces three new parameters into the theory and the charged
fermion masses are related to the fundamental couplings in the theory as
follows:
\begin{eqnarray}
M_u~=~ h \kappa_u + f v_u~+ h'v_u \\  \nonumber
M_d~=~ h \kappa_d + f v_d ~+h'v_d\\  \nonumber
M_\ell~=~ h \kappa_d -3 f v_d -3h'v_d\\  \nonumber
M_{\nu_D}~=~ h \kappa_u -3 f v_u -3h'v_u\\
\end{eqnarray}
Note that the extra contribution compared to Eq. (10) is antisymmetric
which
therefore does not interfere with the mechanism that lead to ${
M}_{\nu,33}$ becoming small as a result of $b-\tau$ convergence. Hence the
natural way that $\theta_A$ became large in the CP conserving case
remains. 

Let us discuss if the new model is still predictive in the neutrino
sector. Of the three new parameters, one is determined by the CP violating
quark phase. the two others are determined by the solar mixing angle and
the solar mass difference squared. Therefore we lose the prediction for
these parameters. However, we can predict in addition to $\theta_A$ (see
above), $\theta_{13}$ and the Dirac phase for the neutrinos. 

\section{Radiative generation of large mixings: another application of
type II seesaw}
As alluded before, type II seesaw liberates the neutrinos from obeying
normal generational hierarchy and instead could easily be quasi-degenerate
in mass. This provides a new mechanism for understanding the large
mixings. The basic idea is that at the seesaw
scale, all mixings angles are small.
 Since the observed neutrino mixings are the weak scale
observables, one must extrapolate\cite{babu1} the seesaw scale mass
matrices to the weak scale and recalculate the mixing angles.
The extrapolation formula is ${ M}_{\nu}(M_Z)~=~ {\bf I}{ M_{\nu}}
(v_R) {\bf I}$
where ${\bf I}_{\alpha \alpha}~=~
\left(1-\frac{h^2_{\alpha}}{16\pi^2}\right)$.
Note that since $h_{\alpha}= \sqrt{2}m_{\alpha}/v_{wk}$ ($\alpha$ being
the charged lepton index), in the extrapolation only the $\tau$-lepton
makes a difference. In the MSSM, this increases the ${ M}_{\tau\tau}$
entry of the neutrino mass matrix and essentially leaves the others
unchanged. It was shown\cite{balaji} that if the muon and the tau neutrinos are
nearly degenerate but not degenerate enough in mass at the seesaw scale,
the radiative corrections can become large enough so that at the weak
scale the two diagonal elements of ${ M}_{\nu}$ 
 become much more degenerate. This leads to an enhancement of the
mixing angle to become almost maximal value.
This can also be seen from the renormalization group equations when they
are written in the mass basis\cite{casas}. Denoting the mixing angles as
$\theta_{ij}$ where $i,j$ stand for generations, the equations are:
\noindent
\begin{eqnarray}
\dsodt&=&-F_{\tau}{c_{23}}^2\left(
-s_{12}U_{\tau1}D_{31}+c_{12}U_{\tau2}D_{32}
\right),\label{eq3}\\
\dstdt&=&-F_{\tau}c_{23}{c_{13}}^2\left(
c_{12}U_{\tau1}D_{31}+s_{12}U_{\tau2}D_{32}
\right),\label{eq4}\\
\dsthdt&=&-F_{\tau}c_{12}\left(c_{23}s_{13}s_{12}U_{\tau1}
D_{31}-c_{23}s_{13}c_{12}U_{\tau2}D_{32}\right.\nonumber \\
&&\left.+U_{\tau1}U_{\tau2}D_{21}\right).\label{eq5}\end{eqnarray}
\noindent
where $D_{ij}={\left(m_i+m_j)\right)/\left(m_i-m_j\right)}$ and
$U_{\tau 1,2,3}$ are functions of the neutrino mixings angles. The
presence
of $(m_i-m_j)$ in the denominator makes it clear that as $m_i\simeq m_j$,
that particular coefficient becomes large and as we extrapolate from the
GUT scale to the weak scale, small mixing angles at GUT scale become large
at the weak scale.
It has been shown recently that indeed such a mechanism for understanding
large mixings can
work for three generations\cite{par}. It was shown that if we identify
the seesaw scale
neutrino mixing angles with the corresponding quark mixings and assume
quasi-degenerate neutrinos, the weak scale solar
and atmospheric angles get magnified to the desired level while due to the
extreme smallness of $V_{ub}$, the magnified value of $U_{e3}$ remains
within its present upper limit. Such a situation can naturally arise in a
parity symmetric model with quark-lepton unification. In figure 2, we show
the evolution of the mixing angles to the weak scale.
 \begin{figure}
\epsfxsize=8.5cm
\epsfbox{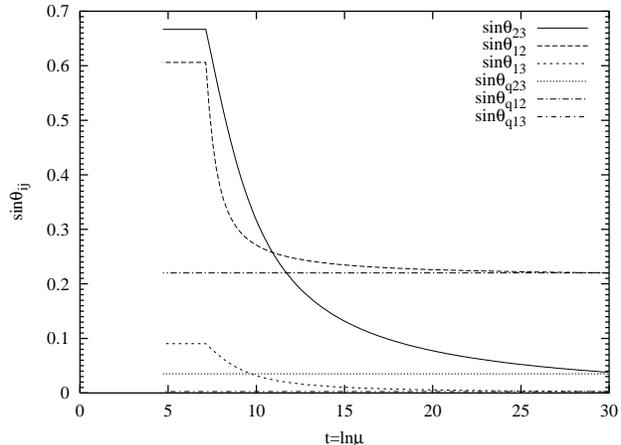}
\caption{Radiative magnification of small quark-like neutrino mixings at
the see-saw scale to bilarge values at low energies. The solid, dashed and
dotted lines represent
$\sin\theta_{23}$, $\sin\theta_{13}$, and $\sin\theta_{12}$,
respectively.}
\label{fig1}
\end{figure}
A requirement for this scenario to work is that the common mass of
neutrinos must be larger than $0.1$ eV, a result that can be tested in
neutrinoless double beta experiments.

\section{Other realizations of seesaw}
As we saw from the previous discussion, the conventional seesaw mechanism
requires rather high scale for the B-L symmetry breaking and the
corresponding right handed neutrino mass (of order $ 10^{15}$
GeV). There is however no way at
present to know what the scale of B-L symmetry breaking is. There are
for example models bases on string compactification\cite{langacker} where
the $B-L$ scale is quite possibly in the TeV range. In this case small
neutrino mass can be implemented by a double seesaw mechanism suggested in
Ref.[21]. The idea is to take a right handed neutrino $N$ whose
Majorana mass is forbidden by some symmetry and a
singlet neutrino $S$ which has extra quantum numbers which prevent it from
coupling to the left handed neutrino but which is allowed to couple
to the right handed neutrino. One can then write a three by three
neutrino mass matrix in the basis $(\nu, N, S)$ of the form:
\begin{eqnarray}
{ M}~=~\left(\begin{array}{ccc} 0 & m_D & 0 \\ m_D & 0 & M \\ 0 & M
& \mu\end{array}\right)
\end{eqnarray}
For the case $\mu \ll M \approx M_{B-L}$, (where $M_{B-L}$ is the $B-L$
breaking scale), this matrix has one light and two heavy states. The
lightest eigenvalue is given by $m_\nu\sim m_d M^{-1}\mu M^{-1}
m_D$. There is a double suppression by the heavy mass compared to the
usual seesaw mechanism and hence the name double seesaw.
A generalization of this mechanism to the case of three generations is
straightforward. One important point here is that to keep $\mu\sim m_D$,
one also needs some additional gauge symmetries, which often are a part
of the string models. In fact this mechanism is sometimes invoked in
string models with TeV scale $Z'$ to understand neutrino
masses\cite{langacker}. 

 It has recently been noted\cite{barr} that if there is parity
symmetry in these models, the $13$ and $31$ entries of the above neutrino 
mass matrix get filled by a small seesaw suppressed entry. This has
interesting applications in neutrino model building.

There are also other alternatives to seesaw,
\cite{alt} discussed in literature, that we do not discuss.

\section{Conclusion}
In summary, the seesaw mechanism is by far the simplest and most appealing
way to understand neutrino masses. It not only improves the aesthetic
appeal of the standard model by restoring quark-lepton symmetry but it
also makes weak interactions asymptotically parity conserving. Further
more it connects neutrino masses with the hypothesis of grand
unification. In this talk I have discussed three
different realizations of this mechanism : type I, type II and double
seesaw. I have also noted how using the type II seesaw mechanism, one can
have simple understanding of large neutrino mixings among neutrinos. A
particularly interesting framework is provided by the minimal SO(10) model
with {\bf 126} Higgs fields, which provides a very predictive framework
for neutrinos.

 This work is supported by the National Science Foundation
Grant No. PHY-0354401. I would also like thank the organizers of the
SEESAW25 conference for the invitation.

Note added: After the Seesaw25 conference in Paris, it came to the
attention of the author as well as others in the community that there was
an early 1977 paper by P. Minkowski (Phys. Lett. {\bf B 67}, 421 (1977))
where the seesaw (type I) formula was also discussed.

\end{document}